\documentclass[aps,prc,twocolumn,superscriptaddress,preprintnumbers]{revtex4-2}
\usepackage[utf8]{inputenc}
\usepackage{physics}
\usepackage{slashed}
\usepackage{amssymb}
\usepackage{cancel}
\usepackage{xcolor}
\usepackage[inline]{enumitem}
\usepackage{comment}

\newcommand\numberthis{\addtocounter{equation}{1}\tag{\theequation}}
\def\ba#1\ea{\begin{align*}#1\numberthis{}\end{align*}}
\newcommand{\be}{\begin{equation}}
\newcommand{\ee}{\end{equation}}

\newcommand{\reb}[1]{[{#1}]}
\newcommand{\rob}[1]{\left({#1}\right)}

\newcommand{\f}[2]{\frac{#1}{#2}}
\newcommand{\bs}[1]{\boldsymbol{#1}}

\newcommand{\de}{\delta}

\usepackage[sep=3pt, offset=1.5em]{simpler-wick}
\usepackage{mathtools}

\newcommand{\q}[1]{\hat{Q}_{{#1}}}

\newcommand{\p}[1]{\hat{P}_{{#1}}}
\newcommand{\opt}[1]{\hat{T}_{{#1}}}

\newcommand{\opty}[1]{\hat{T}^\infty_{{#1}}}

\newcommand{\vop}{\hat{V}}

\newcommand{\gop}{\hat{G}_0}
\newcommand{\calO}{{\cal O}}

\newcommand{\aop}{{\cal\hat{A}}}

\newcommand{\avec}{{\bs{a}}}

\newcommand{\dvec}{{\bs{d}}}
\newcommand{\ivec}{{\bs{i}}}

\newcommand{\kvec}{\bs{k}}

\newcommand{\rvec}{{\bs{r}}}
\newcommand{\svec}{{\bs{s}}}
\newcommand{\pvec}{{\bs{p}}}

\newcommand{\Kvec}{\bs{K}}

\newcommand{\nph}[1]{$#1p#1h$}
\def\({\left(} 
\def\){\right)}
\def\[{\left[} 
\def\]{\right]}

\def\kt{\rangle}
\begin{document}
\preprint{ LA-UR-23-24947}
%
%
\title{The asymptotic behaviour of the many-body coupled cluster amplitudes}

\author{Saar Beck}
\affiliation{The Racah Institute of Physics, The Hebrew University, 
Jerusalem 9190401, Israel}
\author{Ronen Weiss}
\affiliation{Theoretical Division, Los Alamos National Laboratory, Los Alamos, New Mexico 87545, USA}
\author{Nir Barnea}
\affiliation{The Racah Institute of Physics, The Hebrew University, 
Jerusalem 9190401, Israel}
\date{\today}
%
%
\begin{abstract}
  We analyze the asymptotic behaviour of the coupled cluster many-body wave-function 
  in the limit of  highly excited two- and three-particle states.
  We find that in this limit the different coupled cluster amplitudes 
  exhibit a recurring 
  behaviour, factorizing into a common asymptotic two- or three-body term.
  These asymptotic terms depend on the potential and in general are system specific.
  We also suggest that the knowledge of the asymptotic behaviour can potentially help solving the 
  coupled cluster equations in a more efficient way.
\end{abstract}
\maketitle
%
%
\section{Introduction}
Since its inception more than half a century 
ago, Coupled Cluster (CC) theory  \cite{coester58,coester60,cizek66,cizek71} 
had become an
indispensable tool in computational quantum chemistry
\cite{Bartlett07rev} and in nuclear physics \cite{hagen14rev}, allowing for a reliable yet
computationally affordable approximation of the many-body wave-function. 
The advantage of the CC method stems from explicit introduction of  cluster amplitudes that describe $n$-body correlations and
are summed to all orders.
This property of the CC method also makes it
size-extensive and size-consistent 
\cite{Bartlett78ext,Bartlett81ext,Nooijen05ext},
and a natural framework for studying the role and structure of correlations within the many-body system.

Recently, the study of short range correlations (SRCs) has attracted much attention in different and independent disciplines such as atomic \cite{SRCsCoulomb2013}, 
molecular \cite{Tan08a,Tan08b,Tan08c,Braaten12},  and
nuclear physics \cite{Atti:2015eda,Hen:2016kwk}. Motivated by this progress
we study here the manifestation of SRCs within a generic CC fermionic many-body wave-function.
More specifically, we study the properties of the $n$-body CC amplitudes
in the limit of highly excited 2-body and 3-body states.

Assuming a Hamiltonian containing only 2-body physical interaction, and under some constraints regarding the nature of this interaction, we show that in this limit of highly excited 2-body (3-body) states, the $n$-body CC amplitudes factorize into a product of 2-body (3-body) correlation operator
times a residual term, and
as a result, each $n$-body CC amplitude assumes asymptotically the same behaviour as the 2-body or 3-body CC amplitude.

These observations are closely 
related to the factorization ansatz \cite{Tan08a,Braaten12,Werner12,Weiss14,Weiss:2015mba,Weiss_2016,Weiss17_CoupledChannels}
used in the study of SRCs,  which states that if a pair of
nucleons, say $ij$, has high relative momentum $\kvec_{ij}$ but low center of mass momentum $\Kvec_{ij}$ then the many-body wave-function can be written as a product of  the zero-energy eigenstate of the 2-body
Schr{\"o}dinger equation $\varphi$ and a regular function $A$ that describes
the residual system, i.e.
\begin{equation}
   \Psi(\kvec_1,\kvec_2,...,\kvec_A)
   \xrightarrow[k_{ij}\rightarrow \infty]{}
     \varphi(\kvec_{ij})
      A\left(\Kvec_{ij},\{\kvec_n\}_{n\neq i,j}\right).
\end{equation}
Here $\kvec_n$ stands for momentum of particle $n$.

The factorization ansatz was proven to be correct for systems with zero-range interaction
\cite{Tan08a,Braaten12}, i.e.  for a system where the interaction range is much smaller than the average inter particle distance and the scattering length.

In a recent paper \cite{BWB23A} we have utilized the CC theory and 
showed that asymptotically, for nuclear matter, the 2-body cluster amplitude takes a simple form related to the 2-body zero-energy Schr{\"o}dinger equation eigenstate. A similar situation happens also for the 3-body cluster amplitude. The observations presented here generalize these results for an arbitrary $n$-body amplitude. As such, they may also serve as 
a starting point for approximating or solving for these
amplitudes.

The paper is organized as follows. In section \ref{sec:cc} we give a brief recap of CC theory and notation.
Then, for the sake of clarity, in section \ref{sec:main_res} we detail the main 
assumptions and results of this paper without any derivations.
The derivations are given in section \ref{sec:2b_reduction} for the factorization of the 2-body asymptotics, and in section \ref{sec:3b_reduction} for the 3-body case.
Finally, in section \ref{sec:cc_eq_asym} we demonstrate how the knowledge of the asymptotic behaviour can be used to eliminate it from the CC amplitudes.

\section{Recap of coupled cluster theory}\label{sec:cc}

Considering a system of $A$ identical fermions, such as nucleons or atoms, interacting via a 2-body potential, we may write the Hamiltonian of the system in the form
\begin{align}
   \hat{H} &\equiv \hat{H}_0+\hat{U}+\vop 
   \cr &= 
   \sum_{r} \epsilon_r \bs{r}^\dagger\bs{r}
   +\sum_{rs} U^r_s \bs{r}^\dagger\bs{s}
    +\f{1}{4}\sum_{rsr's'}V^{rs}_{r's'}\bs{r}^\dagger\bs{s}^\dagger\svec'\rvec',
\end{align}
where $\hat{H}_0$ is the 1-body, ``zero-order" or unperturbed Hamiltonian (not necessarily the free Hamiltonian),
$\hat{U}$ is a 1-body potential, and 
$\vop$ is a 2-body interaction. 
The 1-body potential $\hat{U}$ has two possible sources: 
\begin{enumerate*}[label=(\alph*)]
\item an external, physical, potential, or
\item a residual interaction emerging from the definition of $\hat{H}_0$. 
\end{enumerate*}
Noting that the first source might affect the asymptotic behaviour of the wave-function,
to simplify the discussion, in the following we will limit our attention 
to the second.

The operators $\rvec^\dagger,\svec,\ldots$ are the usual fermionic creation and annihilation operators, 
corresponding to the single particle eigenstates $\ket{r},\ket{s},\ldots$ of the unperturbed Hamiltonian, 
i.e. $\hat{H}_{0} \ket{{s}} = \epsilon_s \ket{{s}}$.
They obey the anti-commutation relations
\be \label{eq:anticom}
   \{\rvec,\svec\}=0,
   \hspace{2em}
   \{\rvec^\dagger,\svec^\dagger\}=0,
   \hspace{2em}
   \{\rvec^\dagger,\svec\} = \de_{rs}.
\ee

The Slater-determinant state $\ket{\Phi_0}$,
composed of $A$ single particle states, is often a good initial guess for the system's ground state.
The exact wave function $|\Psi\rangle$ is, however, a linear combination of all such Slater determinants. These determinants can be organized in a systematic way, by considering first the determinants obtained by replacing a single state occupied in $\ket{\Phi_0}$ (a ``hole'' state) with a state not occupied in $\ket{\Phi_0}$ (a ``particle'' state), 
then replacing two such states, and so on.
Using the convention of Ref. \cite{shavit2009}, the letters $i,j,\ldots $ denote ``hole'' states, and the letters $a,b,\ldots$ denote ``particle'' states. The letters $r,s,\ldots$ denote both states. It follows that,
\ba
    \bs{i}^\dagger \ket{\Phi_0} = 0\;,
    \quad \text{and} \quad
    \bs{a} \ket{\Phi_0} =0.
\ea
In CC theory, the interacting many-body wave-function $|\Psi \rangle$ is written as
\be\label{Psi_CC}
  \ket{\Psi} = e^{\opt{}} \ket{\Phi_0},
  \quad\text{where}\quad
  \opt{} = \sum_n \opt{n},
\ee
and
\be
  \opt{n} = \f{1}{n!^2}\sum_{a_1 \ldots a_n, i_1 \ldots i_n}
    t^{a_1 a_2\ldots a_n}_{i_1 i_2 \ldots i_n} 
       \bs{a}_1^\dagger\bs{a}_2^\dagger\cdots \avec_n^\dagger
       \ivec_n \cdots \bs{i}_2\bs{i}_1 
\ee
is the $n$-particle, $n$-hole (\nph{n}) cluster operator.
As such, the CC amplitude $t^{a_1\cdots a_n}_{i_1\cdots i_n}$ describes a
correlated $n$-body excitation from an $n$-hole state.  

If the Slater determinant $\ket{\Phi_0}$ is non degenerated, the energy $E$ and the CC amplitudes $t^{a_1 a_2\ldots a_n}_{i_1 i_2 \ldots i_n}$, can be determined by projecting the Schr{\"o}dinger equation on an \nph{n} states such as 
$\ket{\Phi^{a_1a_2\cdots a_n}_{i_1i_2\cdots i_n}}\equiv
\avec^\dagger_1\avec^\dagger_2\cdots\avec^\dagger_n\ivec_n\cdots \ivec_2\ivec_1 \ket{\Phi_0}$ \cite{HM99}. The resulting equations are
\be \label{eq:cc_eq_c0}
   \bra{\Phi_0}
    e^{-\opt{}}\hat{H}e^{\opt{}}\ket{\Phi_0} = E \;,
\ee
and
\be \label{eq:cc_eq_cn}
   \bra{\Phi^{a_1a_2\cdots a_n}_{i_1i_2\cdots i_n}}
    e^{-\opt{}}\hat{H}e^{\opt{}}\ket{\Phi_0} = 0 \;.
\ee
Eq. \eqref{eq:cc_eq_cn} holds for any $n>0$ \nph{n} state, and can be used
to determine the amplitude $t^{a_1 a_2\ldots a_n}_{i_1 i_2 \ldots i_n}$.

In practice the exponents appearing in \eqref{eq:cc_eq_c0},\eqref{eq:cc_eq_cn} are expanded utilizing the Hausdorff expansion
\ba\label{eq:hausdorff}
  e^{-\opt{}}&\hat{H} e^{\opt{}} = \hat{H} + [\hat{H},\opt{}\,]
  + \f{1}{2!}[[\hat{H},\opt{}\,],\opt{}\,]
  \cr &
   +\f{1}{3!}[[[\hat{H},\opt{}\,],\opt{}\,],\opt{}\,]
   +\f{1}{4!}[[[[\hat{H},\opt{}\,],\opt{}\,],\opt{}\,],\opt{}\,],
\ea
which terminates after fourfold commutations as the Hamiltonian includes only 1-body and 2-body terms.

Inserting \eqref{eq:hausdorff} into \eqref{eq:cc_eq_cn}, we note that
the first term on the right-hand side appears only in the \nph{1} and \nph{2} equations, and that 
$\hat{H}_0$ contributes only to the second term yielding
\be
   \bra{\Phi^{a_1a_2\cdots a_n}_{i_1i_2\cdots i_n}}
       [\hat{H}_0,\opt{}]\ket{\Phi_0} = 
       E^{a_1a_2\cdots a_n}_{i_1i_2\cdots i_n}t^{a_1a_2\cdots a_n}_{i_1i_2\cdots i_n} \;,
\ee
where 
$ E^{a_1a_2\cdots a_n}_{i_1i_2\cdots i_n} = \sum_{j=1}^n\epsilon_{a_j}
 -\sum_{j=1}^n\epsilon_{i_j}$.

For completeness, we also present here the full two-body CC equation
neglecting the \nph{1} cluster operator $\opt{1}$ and the 1-body residual potential $\hat{U}$,
\ba\label{eq:2b_3b_cc_eq}
   \bra{\Phi_{i_1i_2}^{a_1a_2}}
  \vop&+[\hat{H}_{0},\opt{2}]+ [\vop,\opt{2}]
   +\f{1}{2}[[\vop,\opt{2}],\opt{2}]
   \cr &
    + [\vop,\opt{3}]+[\vop,\opt{4}]
   \ket{\Phi_0}=0.
\ea

%
%
\section{Assumptions and Main Results}\label{sec:main_res}

Before diving into a detailed analysis of the CC equations, in this section
we present a short summary of the asymptotic properties of the CC amplitudes, and of the assumptions made in order to derive them.

Some of these assumptions are common to many CC applications. These are
the following assumptions:
\begin{enumerate}[label=(\itshape\roman*)]
\item The analysis is limited to closed shell systems, i.e. systems where $\hat{H}_0$ has non degenerated ground state. 
\item The eigenstates of $\hat{H}_0$ are discrete and ordered by their energies.
\item The discussion is limited to systems of fermions interacting via
2-body potential with no physical 1-body or 3-body forces. 
\end{enumerate}
The last assumption rules out, for example, the electrons in the atom which interact with the coulomb field of the static nucleus.

Then there are assumptions that are more specific to our analysis,
and are related to the nature of the interaction. 
To make them more tangible it is convenient to work in momentum space, 
setting the system in a box of size $L^3$
with periodic boundary conditions. 
In this setting $\hat{H}_0$ is the kinetic energy operator, and a state $s$ stands for the particle's momentum $\pvec_s$, and its internal degrees of freedom.
These assumptions are:
\begin{enumerate}[label=(\itshape\roman*)]
\item The matrix elements of the 2-body interaction are bounded, 
i.e. there is a 
bound $M$ such that for any $r,s,r',s'$ the relation 
$|V^{r s}_{r' s'}|=|\bra{r s} \vop \ket{r' s'}| \leq M$ holds. 
\item Momentum conservation - 
The matrix elements of the potential
$V^{r s}_{r's'}$ vanish unless $\pvec_r+\pvec_s=\pvec_{r'}+\pvec_{s'}$.
Consequently $t^{d_1 \cdots d_n}_{i_1\cdots i_n}$ vanish unless 
$\sum_j \pvec_{d_j}=\sum_j \pvec_{i_j}$.
\end{enumerate}

As we shall see in the following sections, the momentum conservation condition plays a major role in our analysis. However, it seems that we can do with a weaker condition. Namely, that in the asymptotic limit $a\to\infty$ the potential matrix elements $V^{a b}_{i j}$, or $V^{a i}_{b j}$ vanish unless $b\to\infty$ and $\epsilon_b\approx \epsilon_a$. 
Nevertheless, when referring
to highly excited 2-particle states we consider the situation  
where the 2 states have similar energies but
the potential matrix element does not vanish, i.e. $V^{a_1a_2}_{i_1i_2}\neq 0$.  
This condition implies that the center of mass momentum $\pvec_{a_1}+\pvec_{a_2}$
is limited by the Fermi surface. A similar condition holds also for the 3-particle case.

We also note that our particular choice of $\hat{H}_0$ implies that
$\hat{U}=0$, and that our choice of
basis states implies that $\opt{1}=0$ \cite{CCNM2013}. 

Under these assumptions we found out that the CC amplitude $t^{d_1\cdots d_n}_{i_1\cdots i_n}$, describing the $n$-body correlation excitation, takes on a particular form in the limit of highly excited 2, or 3-particle states.
Specifically, in the limit $a_1,a_2\to \infty$
\be
   t^{a_1a_2d_3\cdots d_n}_{i_1\cdots  i_n} \longrightarrow
   \f{1}{2}\tau^{a_1a_2}_{r_1r_2}
   L^{r_1r_2;d_3\cdots d_n}_{i_1\cdots i_n} 
\ee 
where $\hat{\tau}_2$ is a 2-body operator, given by
\be
   \hat{\tau}_2 = \f{1}{1-\q{2}\gop\rob{0}\vop}\q{2}\gop\rob{0}\vop\;,
\ee
and $L^{r_1r_2;d_3\cdots d_n}_{i_1\cdots i_n}$ is a low energy $n$-body amplitude
that depends on the CC amplitudes $\opt{2},\ldots,\opt{n-1}$, but is
independent of $a_1,a_2$.
Here
$\gop\rob{E}=\f{1}{E+i\varepsilon-\hat{H}_0}$ is the Green's function
and $\q{n}$ is the $n$-body projection operator into the particle subspace.
$\hat{\tau}_2$ is closely related to the asymptotics of the 2-body cluster operator $\opty{2}$ introduced in \cite{BWB23A}
and admits $\tau^{a_1a_2}_{i_1i_2}= (t^\infty)^{a_1a_2}_{i_1i_2}$,
that is, the $n$-body CC amplitude
has a similar behaviour
to the  2-body
CC amplitude of the pair times a residual term.

A similar factorization arises when a triplet is considered. 
When $a_1,a_2,a_3\to \infty$ the CC
amplitude factorizes into
\be
   t^{a_1a_2a_3d_4\cdots d_n}_{i_1\cdots  i_n} \longrightarrow
   \f{1}{6}\tau^{a_1a_2a_3}_{r_1r_2r_3}
   L^{r_1r_2r_3;d_4\cdots  d_n}_{i_1\cdots i_n} 
\ee 
where $\hat{\tau}_3$ is given by
\be
   \hat{\tau}_3=\f{1}{1-\q{3}\gop\rob{0}\vop}\q{3}\gop\rob{0}\vop
   \hat{\tau}_2\;,
\ee
and $L^{r_1r_2r_3;d_4\cdots d_n}_{i_1\cdots i_n}$ is the low energy $n$-body amplitude
associated with 3-body correlations.
Also $\hat{\tau}_3$ is related to the 3-body CC amplitude asymptotics \cite{BWB23A}
and admits the relation 
$\tau^{a_1a_2a_3}_{i_1i_2i_3}= (t^\infty)^{a_1a_2a_3}_{i_1i_2i_3}$.
%
%
\section{The 2-body factorization of the coupled cluster amplitudes}
\label{sec:2b_reduction}
To compute the structure of the $n$-body CC amplitude $\opt{n}$ in the limit of 2 highly excited
particle states, say $a_1,a_2$, 
we need to project Eq. \eqref{eq:cc_eq_cn} on the \nph{n} state 
$|\Phi^{a_1a_2d_3\ldots d_n}_{i_1\ldots i_n}\kt = 
 \avec_1^\dagger\avec_2^\dagger \dvec_3^\dagger \cdots \dvec_n^\dagger
 \ivec_n \cdots \ivec_1\ket{\Phi_0}$. 
We are interested in the case where $d_j$ indicates particle state not too far from the Fermi level and $a_1,a_2\to\infty$ in a fashion that $V^{a_1a_2}_{i_1i_2}\neq 0$. We also note that in order for the wave-function to be properly normalized the amplitude $t^{a_1a_2d_3\cdots d_n}_{i_1\cdots i_n}$ must vanish when $a_1,a_2\to\infty$.

However, before we start analyzing the full CC equations, let us begin with a simplified case, known as CC with doubles - CCD, where only 2-body correlations are included in the CC expansion, i.e. $\opt{}=\opt{2}$. In this case Eq. \eqref{eq:cc_eq_cn} (or Eq. \eqref{eq:2b_3b_cc_eq}) takes the form
\ba\label{ccd_eq}
   0 &= V^{a_1a_2}_{i_1i_2}+E^{a_1a_2}_{i_1i_2}t^{a_1a_2}_{i_1i_2}
   +\f{1}{2}V^{a_1a_2}_{c_1c_2}t^{c_1c_2}_{i_1i_2}
   \cr &
  +\aop^{a_1a_2}V^{a_1j_1}_{j_1c_1}t^{a_2c_1}_{i_1i_2}   
   +\aop^{a_1a_2}_{i_1i_2}V^{a_1j_1}_{i_1c_1}t^{a_2c_1}_{i_2j_1}
   \cr &
   +\f{1}{2}V^{j_1j_2}_{i_1i_2}t^{a_1a_2}_{j_1j_2}
  +\aop_{i_1i_2}V^{j_1j_2}_{i_1j_2}t^{a_1a_2}_{i_2j_1}
   \cr &
   +\aop_{i_1i_2}V^{j_1j_2}_{c_1c_2}t^{a_1c_2}_{i_1j_2}t^{c_1a_2}_{j_1i_2}
   -\f{1}{2}\aop^{a_1a_2}V^{j_1j_2}_{c_1c_2}t^{a_1c_2}_{j_1j_2}t^{c_1a_2}_{i_1i_2}
   \cr &
   +\f{1}{4}V^{j_1j_2}_{c_1c_2}t^{c_1c_2}_{i_1i_2}t^{a_1a_2}_{j_1j_2}
   -\f{1}{2}\aop_{i_1i_2}V^{j_1j_2}_{c_1c_2}t^{c_1c_2}_{i_1j_2}t^{a_1a_2}_{j_1i_2}
   .
\ea
Here $\aop^{d_1\cdots d_n}_{j_1\cdots j_n}$ is the anti-symmetrization operator, acting separately on the upper and lower indices, and we also use the convention that when the same state appears as both upper and lower index, summation should be understood. 

Inspecting Eq. \eqref{ccd_eq} we note that the states $a_1,a_2$ may be contracted 
\begin{enumerate*}[label=(\alph*)]
\item both with the potential $\vop$ (1st and 3rd terms on the first line), 
\item one with $\vop$ and one with $\opt{2}$ (2nd line), or
\item both with the CC amplitudes
(lines 3-5 and the second term in the first line). 
\end{enumerate*}
Due to the momentum conservation property all terms but case (a) are proportional to sums over $j_1,j_2$ of the amplitudes $t^{a_1a_2}_{j_1j_2}$ where $\epsilon_{a_1}\approx \epsilon_{a_2}$.

Assuming that the rate of convergence to zero of the CC amplitudes $t^{a_1a_2}_{j_1j_2}$ in the limit $a_1,a_2\to\infty$ is independent of the hole states,
we may conclude that the terms
on the 2nd and 4th lines of \eqref{ccd_eq}, such as 
$V^{a_1j_1}_{j_1c_1}t^{a_2c_1}_{i_1i_2}$, 
must be bounded by $M' |t^{a_1a_2}_{i_1 i_2}|$, or $M'' |t^{a_1a_2}_{i_1 i_2}|^2$,
and as such much smaller asymptotically than the $[\hat{H}_0,\opt{2}]$ term $E^{a_1a_2}_{i_1i_2}t^{a_1a_2}_{i_1i_2}$.
The terms appearing on the 3rd and 5th lines of \eqref{ccd_eq} can be regarded
as matrices connecting 2-holes to 2-holes, acting on the asymptotic amplitudes $t^{a_1a_2}_{j_1j_2}$ or $t^{a_1a_2}_{i_1j_2}$. As these matrices are independent of $a_1,a_2$ these terms should have the same asymptotic behaviour as $t^{a_1a_2}_{i_1i_2}$, and as a result for large enough particle states they also
should be negligible in comparison to the $\hat{H}_0$ term.

Summing up, the leading terms that should be retained in Eq. \eqref{ccd_eq},
when contracted with $a_1,a_2\to\infty$ are only those appearing in the first row.
Relatively, all other terms are vanishingly small.
In this limit $a_1,a_2\to\infty$ we can use the approximation
$E^{a_1a_2}_{i_1i_2}\approx E^{a_1a_2}$ and write down the asymptotic equation
\be\label{eq_t2_asym}
  0 = V^{a_1a_2}_{i_1i_2}+E^{a_1a_2}_{}t^{a_1a_2}_{i_1i_2}
     +\f{1}{2}V^{a_1a_2}_{c_1c_2}t^{c_1c_2}_{i_1i_2}
  \;.
\ee
Following Ref. \cite{BWB23A} we denote the solution of this equation 
as $\opty{2}$, and note that
$ \opty{2}=\hat{\tau}_2\hat{P}_2 $,
where $\hat{P}_n$ is the $n$-body projection operator into the holes subspace.
  By construction, $\opty{2}$ is the asymptotic 2-body CC amplitude 
$t^{a_1a_2}_{i_1i_2}\to \rob{t^\infty}^{a_1a_2}_{i_1i_2}$.
From Eq. \eqref{eq_t2_asym} we can also conclude that asymptotically
$t^{a_1a_2}_{i_1i_2} \propto V^{a_1a_2}_{r_1r_2}/E^{a_1a_2}$, i.e. 
in the limit $a_1,a_2\to\infty$ the CC amplitude falls off much faster than the potential. 

To generalize this result for an arbitrary $n$-body amplitude $\opt{n}$,
we divide the different terms appearing in the $n$-body CC equation \eqref{eq:cc_eq_cn} into three groups: 
\begin{enumerate*}[label=(\itshape\roman*)]
\item The kinetic energy term $E^{a_1a_2d_3\cdots d_n}_{i_1\cdots i_n}t^{a_1a_2d_3\cdots d_n}_{i_1\cdots i_n}$.
\item The terms such as  $V^{a_1a_2}_{c_1c_2}t^{c_1c_2d_3\cdots d_n}_{i_1\cdots i_n}$
where the high momentum indices $a_1,a_2$ are contracted with the potential,
and
\item All the remaining terms, that collectively can be written as
$V^{s_1s_2}_{r_1r_2}R^{r_1r_2;a_1a_2d_3\cdots d_n}_{s_1s_2;i_1\cdots i_n}$.
\end{enumerate*}
We note that due to momentum conservation, each term in 
$\hat{R}_{s_1s_2}^{r_1r_2}$ has at least one CC amplitude contracted with 2
high momentum indices that are equivalent to $a_1,a_2$.

Now, to proceed, we order the various CC amplitudes $t^{a_1a_2d_3\cdots d_k}_{i_1\cdots i_k}$ by the rate in which they converge to zero in the limit $a_1,a_2\to\infty$.
Assuming that such an order exists, we pick the CC amplitude  
with the slowest convergence rate, say $t^{a_1a_2d_3\cdots d_n}_{i_1\cdots i_n}$. 
As the kinetic energy diverges, it is evident
that (at least) for this specific amplitude the kinetic energy term
$E^{a_1a_2d_3\cdots d_n}_{i_1\cdots i_n}t^{a_1a_2d_3\cdots d_n}_{i_1\cdots i_n}$ must be much
larger than the residual term, i.e.
$E^{a_1a_2d_3\cdots d_n}_{i_1\cdots i_n}t^{a_1a_2d_3\cdots d_n}_{i_1\cdots i_n}\gg V^{s_1s_2}_{r_1r_2}R^{r_1r_2;a_1a_2d_3\cdots d_n}_{s_1s_2;i_1\cdots i_n}$.
Hence, in the asymptotic CC equation we can ignore this term, 
and consider only the kinetic energy term
 and the terms where both $a_1,a_2$ are contracted with the potential.

In view of this discussion, the commutators
$\f{1}{3!}[[[\hat{H},\opt{}],\opt{}],\opt{}]$,
$\f{1}{4!}[[[[\hat{H},\opt{}],\opt{}],\opt{}],\opt{}]$ 
can be neglected, and the leading terms coming from the commutators
$[\hat{H},\opt{}]$, $\f{1}{2!}[[\hat{H},\opt{}],\opt{}]$ 
are of the form
\be
   V^{a_1a_2}_{d_1d_2}t^{d_1\cdots d_n}_{i_1\cdots i_n},
   \quad V^{a_1a_2}_{i_1d_2}t^{d_2\cdots d_n}_{i_2\cdots i_n}\;+\text{permutations}
\ee
and
\be
   \sum_{l=2}^{n-2}V^{a_1a_2}_{d_2d_{n+1}}t^{d_2\cdots d_{l+1}}_{i_1\cdots i_l}
   t^{d_{l+2}\cdots d_{n+1}}_{i_{l+1}\cdots i_n}\;+\text{permutations}
   \;.
\ee
Collecting all these terms into the asymptotic $n$-body CC amplitude equation ($n>2$) we get
\ba\label{eq:2b_fctr_asym_pre_l}
   t&^{a_1a_2d_3\cdots d_n}_{i_1\cdots i_n} +
   \f{V^{a_1a_2}_{c_1c_2}}{2E^{a_1a_2d_3\cdots d_n}_{i_1\cdots i_n}}
   t^{c_1c_2d_3\cdots d_n}_{i_1\cdots i_n} =
   \\&-
   \f{\aop_{i_1\cdots i_n}}{\rob{n-1}!}
   \rob{\f{V^{a_1a_2}_{i_1c_2}}{E^{a_1a_2d_3\cdots d_n}_{i_1\cdots i_n}}
         t^{c_2d_3\cdots d_n}_{i_2\cdots i_n}}
   \\&-
   \sum_{l=2}^{n-2} B_{nl}
   \f{V^{a_1a_2}_{c_2c_{n+1}}}{2E^{a_1a_2d_3\cdots d_n}_{i_1\cdots i_n}}
   \aop^{d_3\cdots d_{n}}_{i_1\cdots i_n}
   \rob{t^{c_2d_3\cdots d_{l+1}}_{i_1\cdots i_l}t^{d_{l+2}\cdots c_{n+1}}_{i_{l+1}\cdots i_n}}
   \;,
\ea
with
\be
    B_{nl} = \frac{\rob{-1}^n}{l!\rob{l-1}!\rob{n-l}!\rob{n-l-1}!}\;.
\ee
Defining (for $n>2$) the operator 
 \ba\label{eq:2b_fctr_asym_ll}
 \hat{L}^{r_1r_2}_n=
\aop^{r_1r_2}\rob{\wick{\c1{\hat{\bs{r}}_{2}}\c1{\hat{T}}_{n-1}}\hat{\bs{r}}_{1}}+
\rob{\sum_{l=2}^{n-2}\wick{\c1{\hat{\bs{r}}_{2}}\c1{\hat{T}}_l}\wick{\c1{\hat{\bs{r}}_{1}}\c1{\hat{T}}_{n-l}}}
\ea
with matrix elements
\ba\label{ll2_me}
   &L^{r_1r_2;d_3\cdots d_n}_{i_1\cdots i_n} =
   \f{1}{(n-1)!}\aop^{r_1r_2}_{i_1\cdots i_n}\rob{\delta^{r_1}_{i_1}t^{r_2\cdots d_n}_{i_2\cdots i_n}}
   \cr &\hspace{2em}+
   \sum_{l=2}^{n-2} B_{nl}
   \aop^{d_3\cdots d_{n}}_{i_1\cdots i_n}
   \rob{t^{r_1d_3\cdots d_{l+1}}_{i_1\cdots i_l}t^{d_{l+2}\cdots d_{n}r_2}_{i_{l+1}\cdots i_n}}
\;,\ea
we can rewrite Eq. \eqref{eq:2b_fctr_asym_pre_l} as
\ba\label{eq:2b_fctr_asym}
   t^{a_1a_2d_3\cdots d_n}_{i_1\cdots i_n} &+
   \f{V^{a_1a_2}_{c_1c_2}}{2E^{a_1a_2d_3\cdots d_n}_{i_1\cdots i_n}}t^{c_1c_2d_3\cdots d_n}_{i_1\cdots i_n}
   \cr &=
   - \frac{V^{a_1a_2}_{r_1r_2}}{2E^{a_1a_2d_3\cdots d_n}_{i_1\cdots i_n}}
   L^{r_1r_2;d_3\cdots d_n}_{i_1\cdots i_n}
\;.\ea
For small enough clusters and for large enough ${a_1},{a_2}$ we can safely
use the approximation $E^{a_1a_2d_3\cdots d_n}_{i_1\cdots i_n}\approx E^{a_1a_2}$, and rewrite Eq. \eqref{eq:2b_fctr_asym} as
\ba\label{eq:2b_fctr_asym_2}
   t^{a_1a_2d_3\cdots d_n}_{i_1\cdots i_n}+
   \f{V^{a_1a_2}_{c_1c_2}}{2E^{a_1a_2}}&t^{c_1c_2d_3 \cdots d_n}_{i_1\cdots i_n}
   = 
   \cr &
   - \f{V^{a_1a_2}_{r_1r_2}}{2E^{a_1a_2}} L^{r_1r_2;d_3\cdots d_n}_{i_1\cdots i_n}\;.
\ea
Seeking a solution to the asymptotic Eq. \eqref{eq:2b_fctr_asym_2} 
of the form
\be\label{2b_fctr_ansatz}
    t^{a_1a_2d_3\cdots d_n}_{i_1\cdots i_n} = 
                \f{1}{2}\tau^{a_1a_2}_{r_1 r_2} L^{r_1r_2;d_3\cdots d_n}_{i_1\cdots i_n}
                \;,
\ee
we get
\be\label{eq:2b_fctr_asym_4}
   \frac{1}{2}\(\tau^{a_1a_2}_{r_1r_2} +
      \frac{V^{a_1a_2}_{c_1c_2}}{2E^{a_1a_2}}\tau^{c_1 c_2}_{r_1r_2}
   + \f{V^{a_1a_2}_{r_1r_2}}{E^{a_1a_2}}\) L^{r_1r_2;d_3\cdots d_n}_{i_1\cdots i_n}
   = 0 \;.
\ee
It follows that a sufficient condition for the amplitude ansatz \eqref{2b_fctr_ansatz}
to solve Eq. \eqref{eq:2b_fctr_asym_2} is
\be\label{2b_asym_tau_eq}
   \tau^{a_1a_2}_{r_1r_2} +
      \frac{V^{a_1a_2}_{c_1c_2}}{2E^{a_1a_2}}\tau^{c_1 c_2}_{r_1r_2}
   + \f{V^{a_1a_2}_{r_1r_2}}{E^{a_1a_2}} =0 \;.
\ee 
It should be noted that $r_1,r_2$ are parameters in \eqref{2b_asym_tau_eq}
and that the summations on $c_1,c_2$ run over all particle states. 

Eq.  \eqref{2b_asym_tau_eq} can be written in a matrix form
for the non-symmetrized basis as
\be\label{2b_asym_tau_mat_eq}
  \hat{\tau}_2 - \q{2}\gop\rob{0}\vop\q{2}\hat{\tau}_2
   - \q{2}\gop\rob{0}\vop = 0,
\ee
leading to the formal solution
\be\label{2b_asym_tau}
   \hat{\tau}_2 = \frac{1}{1-\q{2}\gop\rob{0}\vop}\q{2}\gop(0)\vop\;.
\ee

Inspecting this formal solution we may conclude that
the worst-converging amplitude behaves asymptotically as
$t^{a_1a_2d_3\ldots d_n}_{i_1\ldots i_n}\propto -\f{V^{a_1a_2}_{r_1r_2}}{E^{a_1a_2}}$. 
It follows that our arguments and derivations
leading to Eqs. \eqref{eq:2b_fctr_asym_pre_l}-\eqref{2b_asym_tau}
are valid in general for any $n$-body CC amplitude $t^{a_1a_2d_3\cdots d_n}_{i_1\cdots i_n}$
since
$V^{s_1s_2}_{r_1r_2}R^{r_1r_2;a_1a_2d_3\cdots d_n}_{s_1s_2;i_1\cdots i_n}$
must converge to zero at least as fast as $\f{V^{a_1a_2}_{r_1r_2}}{E^{a_1a_2}}$ and therefore can be neglected
in comparison to $V^{a_1a_2}_{r_1r_2}L^{r_1r_2;d_3\cdots d_n}_{i_1\cdots i_n}$.

Before concluding this section we would like to add two remarks.
The first is that in \cite{BWB23A} it was found that asymptotically the 2-body cluster amplitude
takes the form
\be\label{opty2}
  \opt{2}\to \opty{2}=\f{1}{1-\q{2}\gop\rob{0}\vop}\q{2}\gop\rob{0}\vop\p{2}.
\ee
Comparing \eqref{opty2} with \eqref{2b_asym_tau} the  relation $\opty{2}=\hat{\tau}_2\p{2}$ between these two operators, pointed out above, 
becomes clear.

The second remark is that for the $n=2$ case
the right form of the operator $\hat{L}^{r_1r_2;}_2$ is
$L^{r_1r_2;}_{i_1i_2}\equiv 
\delta^{r_1}_{i_1}\delta^{r_2}_{i_2}-\delta^{r_2}_{i_1}\delta^{r_1}_{i_2}$.
With this definition Eq. \eqref{eq:2b_fctr_asym_2} coincides, for $n=2$,
with Eq. \eqref{eq_t2_asym}, and Eq. \eqref{2b_fctr_ansatz} becomes the relation $\opty{2}=\hat{\tau}_2\p{2}$.

\section{The 3-body factorization of the coupled cluster amplitudes}\label{sec:3b_reduction}

In order to analyze the properties of the CC amplitudes $\opt{n}$, $n\geq 3$, in the limit of highly excited particle triplet, say $a_1,a_2,a_3$, we need to project the CC equations \eqref{eq:cc_eq_cn} on the \nph{n} state 
$$|\Phi^{a_1a_2a_3d_4\ldots d_n}_{i_1\ldots i_n}\kt = 
 \avec_1^\dagger\avec_2^\dagger \avec_3^\dagger\dvec_4^\dagger \cdots \dvec_n^\dagger
 \ivec_n \cdots \ivec_1\ket{\Phi_0}\;.$$
Here we assume that all other particle states $(d_4\ldots d_n)$ are close to the Fermi level, and in any case fixed when we take the limit $a_1,a_2,a_3\to \infty$.
Now taking this limit we note that due to the momentum conservation property the matrix elements of the potential
$V^{a_1a_2}_{i_1b_3}$ are zero unless
the incoming particle state $b_3$ is close to $a_3$, and the same for
$V^{a_2a_3}_{i_1b_1},V^{a_3a_1}_{i_1 b_2}$.
This condition is equivalent to the
demand that the center of mass momentum of the triplet is of the order of the Fermi momentum, i.e. $\pvec_{a_1}+\pvec_{a_2}+\pvec_{a_3}\approx \calO(p_F)$.

Turning now to analyze the amplitude equation, we recall that in the 2-body case the leading terms were the $[\hat{H}_0,\opt{n}]$ term, and terms where
the highly excited particle states were contracted with the potential.
Having 3 highly excited states and a potential that is a 2-body operator, 
at least one of the excited states must be contracted with a
CC amplitude.

Nevertheless, following similar arguments to those presented in Sec. \ref{sec:2b_reduction}, we can conclude retrospectively that terms such as 
$V^{j_1j_2}_{i_1i_2}t^{a_1a_2a_3d_4\cdots d_l}_{j_1j_2i_3\cdots i_l}$
or
$V^{a_3j_1}_{d_3i_1}t^{a_1a_2d_3d_4\cdots d_l}_{j_1i_2\cdots i_l}$
go to zero faster than the other source terms such as
$V^{a_1a_2}_{i_1d_3}t^{a_3d_3\cdots d_n}_{i_2\cdots i_n}$
by a factor of $\propto\f{1}{E^{a_1a_2a_3}}$.
Therefore, we expect
that in the limit $a_1,a_2,a_3\to\infty $, the leading terms in the $\opt{n}$ CC equation are those where 
only one high particle state is contracted with a CC amplitude while the
two other states are contracted with the potential.

Utilizing this conclusion we can discard all terms
coming from the commutation relations $\f{1}{3!}\reb{\reb{\reb{\hat{H},\opt{}},\opt{}},\opt{}}$,
and 
$\f{1}{4!}\reb{\reb{\reb{\reb{\hat{H},\opt{}},\opt{}},\opt{}},\opt{}}$.
Collecting now the leading terms from the commutators 
$\reb{\hat{H},\opt{}}$, and $\frac{1}{2!}\reb{\reb{\hat{H},\opt{}},\opt{}}$,
the $\opt{n}$ CC equation ($n\geq 3$), in the 3-body asymptotic limit, reads
\ba\label{eq:pre_pre_b3_decomp}
  &t^{a_1a_2a_3d_4\cdots d_n}_{i_1\cdots i_n}+\f{\aop^{a_1a_2a_3}}{2}\Bigg[
  \rob{\f{V^{a_1a_2}_{c_2c_3}}{2E^{a_1a_2a_3d_4\cdots d_n}_{i_1\cdots i_n}}
    t^{a_3c_2c_3d_4\cdots d_n}_{i_1\cdots i_n}}\Biggr]
  \cr &=
  \f{\aop^{a_1a_2a_3}}{2}\Biggl[\f{\aop_{i_1\cdots i_n}}{\rob{n-1}!}\biggl(
    \f{V^{a_1a_2}_{i_1c_3}}{E^{a_1a_2a_3d_4\cdots d_n}_{i_1\cdots i_n}}
    t^{a_3c_3d_4\cdots d_n}_{i_2\cdots i_n}
    \biggr)
   \cr &
   +\sum_{l=2}^{n-2} B'_{nl}\aop^{d_4\cdots d_{n}}_{i_1\cdots i_n}
   \Biggl(
   \frac{V^{a_1a_2}_{c_3c_{n+1}}}{E^{a_1a_2a_3d_4\cdots d_n}_{i_1\cdots i_n}}
   \cr & \hspace{5em}   \times
   \rob{t^{a_3c_3d_4\cdots d_{l+1}}_{i_1\cdots i_l}t^{d_{l+2}\cdots d_nc_{n+1}}_{i_{l+1}\cdots i_n}}
\Biggr)\Biggr]
\;,\ea
where
\be
   B'_{nl} = \f{\rob{-1}^{n}}{l!\rob{l-2}!\rob{n-l}!\rob{n-l-1}!}\;.
\ee
Inspecting Eq. \eqref{eq:pre_pre_b3_decomp} we note that due to the momentum conservation condition the sums $c_3$ over the particle states on the right hand side are restricted to states close to $a_3$. Thus the amplitudes
$t^{a_3c_3d_4\cdots d_n}_{i_2\cdots i_n}$, $t^{a_3c_3\cdots d_{l+1}}_{i_1\cdots i_l}$ are associated with two highly excited particle states and can be factorized according to Eq. \eqref{2b_fctr_ansatz}.
Utilizing this factorization, while using the approximation 
$E^{a_1a_2a_3d_4\cdots d_n}_{i_1\cdots i_n}\approx E^{a_1a_2a_3}$,
Eq. \eqref{eq:pre_pre_b3_decomp} becomes
\ba\label{eq:3_b_almost}
  &t^{a_1a_2a_3d_4\cdots d_n}_{i_1\cdots i_n} 
  + 
  \f{\aop^{a_1a_2a_3}}{2}\Bigg[
    \rob{\f{V^{a_1a_2}_{c_2c_3}}{2E^{a_1a_2a_3}}t^{a_3c_2c_3d_4\cdots d_n}_{i_1\cdots i_n}}
    \Biggr]
  =\cr &
  \f{\aop^{a_1a_2a_3}}{2}\Biggl\{
  \f{V^{a_1a_2}_{r_3c_3}\tau^{c_3a_3}_{r_1r_2}}{2E^{a_1a_2a_3}}
  \Bigg[-\f{\aop_{i_1\cdots i_n}}{\rob{n-1}!}\biggl(
    \de_{i_1}^{r_3}L^{r_1r_2;d_4\cdots d_n}_{i_2\cdots i_n}
    \biggr)
  \cr &
    +\sum_{l=3}^{n-2}B'_{nl}\aop^{d_4\cdots d_{n}}_{i_1\cdots i_n}
    \rob{L^{r_1r_2;d_4\cdots d_{l+1}}_{i_1\cdots i_l}t^{d_{l+2}\cdots d_nr_3}_{i_{l+1}\cdots i_n}}
    \Biggr]
  \cr &
  +B'_{n 2}\aop^{d_4\cdots d_{n}}_{i_1\cdots i_n}
  \Biggl(
  \f{V^{a_1a_2}_{c_3c_{n+1}}}{E^{a_1a_2a_3}}
  \rob{\tau^{a_3c_3}_{i_1i_2}t^{d_{4}\cdots c_{n+1}}_{i_{3}\cdots i_n}}
  \Biggr)\Biggr\}
\;.\ea
Here on the last line we have separated the case $l=2$ from the sum on $l$,
and replaced the 2-body amplitude by its asymptotic value.

Eq. \eqref{eq:3_b_almost} can be rewritten now as
\ba \label{eq:asym_3_decomp}
  &t^{a_1a_2a_3d_4\cdots d_n}_{i_1\cdots i_n}
  + \frac{\aop^{a_1a_2a_3}}{2}\Bigg[
    \rob{\f{V^{a_1a_2}_{c_2c_3}}{2E^{a_1a_2a_3}}
      t^{a_3c_2c_3d_4\cdots d_n}_{i_1\cdots i_n}}
    \Biggr]
  \cr = &
  -\f{\aop^{a_1a_2a_3}_{r_1r_2r_3}}{24}
  \bigg(\f{V^{a_1a_2}_{r_3c_3}\tau^{c_3a_3}_{r_1r_2}}{E^{a_1a_2a_3}}\biggr)
  L^{r_1r_2r_3;d_4\cdots d_n}_{i_1\cdots i_n} \;,
\ea
with
\ba
  &L^{r_1r_2r_3;d_4\cdots d_n}_{i_1\cdots i_n} = 
  \f{\aop_{i_1\cdots i_n}}{\rob{n-1}!}\biggl(
    \delta_{i_1}^{r_3}L^{r_1r_2;d_4\cdots d_n}_{i_2\cdots i_n}
    \biggr)
  \cr &
    -\sum_{l=3}^{n-2}B'_{nl}\aop^{d_4\cdots d_{n}}_{i_1\cdots i_n}
    \rob{L^{r_1r_2;d_4\cdots d_{l+1}}_{i_1\cdots i_l}t^{d_{l+2}\cdots d_nr_3}_{i_{l+1}\cdots i_n}}
  \cr &
  -B'_{n 2}\aop^{d_4\cdots d_{n}}_{i_1\cdots i_n}
  \Biggl(
  \rob{\de^{r_1}_{i_1}\de^{r_2}_{i_2}-\de^{r_1}_{i_2}\de^{r_2}_{i_1}}
  t^{d_{4}\cdots d_{n}r_3}_{i_{3}\cdots i_n}
  \Biggr)
  \cr &=
  \f{\aop_{i_1\cdots i_n}}{\rob{n-1}!}\biggl(
    \delta_{i_1}^{r_3}L^{r_1r_2;d_4\cdots d_n}_{i_2\cdots i_n}
    \biggr)
  \cr &
    -\sum_{l=2}^{n-2}B'_{nl}\aop^{d_4\cdots d_{n}}_{i_1\cdots i_n}
    \rob{L^{r_1r_2;d_4\cdots d_{l+1}}_{i_1\cdots i_l}t^{d_{l+2}\cdots d_nr_3}_{i_{l+1}\cdots i_n}}
    \;.
\ea
These are the matrix elements of the $n$-body operator ($n>3$)
  \ba
  \hat{L}^{r_1r_2r_3}_n&=
  \hat{L}^{r_1r_2}_{n-1}\hat{\bs{r}}_3
  +
  \wick{\c1{\hat{\bs{r}}_3}\c1{\hat{T}}_{n-2}}\hat{\bs{r}}_2\hat{\bs{r}}_1
  +
  \sum_{l=3}^{n-2}\wick{\c1{\hat{\bs{r}}_3}\c1{\hat{T}}_{n-l}}\hat{L}^{r_1r_2}_{l}
  \;,\ea
which by using the definitions of $\hat{L}^{r_1r_2}_n$ can also 
be written as
  \begin{widetext}
      \ba
  \hat{L}^{r_1r_2r_3}_n&=
\f{\aop^{r_1r_2r_3}}{2}\rob{\wick{\c1{\hat{\bs{r}_{3}}}\c1{\hat{T}}_{n-2}}\hat{\bs{r}}_{2}\hat{\bs{r}}_{1}}
+
\f{\aop^{r_1r_2r_3}}{2}\rob{\sum_{l=2}^{n-3}\wick{\c1{\hat{\bs{r}}}_{3}\c1{\hat{T}}_l}\wick{\c1{\hat{\bs{r}}_{2}}\c1{\hat{T}}_{n-l-1}}\hat{\bs{r}}_{1}}
+
  \sum_{k=2}^{n-4}\sum_{m=2}^{n-k-2}\wick{\c1{\hat{\bs{r}}_{3}}\c1{\hat{T}}_k}\wick{\c1{\hat{\bs{r}}_{2}}\c1{\hat{T}}_m}\wick{\c1{\hat{\bs{r}}_{1}}\c1{\hat{T}}_{n-k-m}}
  \;.\ea
\end{widetext}

Guessing now a solution to Eq. \eqref{eq:asym_3_decomp}
of the form
\ba\label{eq:tau3_def}
  t^{a_1a_2a_3d_4\cdots d_n}_{i_1\cdots i_n} = 
  \f{1}{6}\tau^{a_1a_2a_3}_{r_1r_2r_3}L^{r_1r_2r_3;d_4\cdots d_n}_{i_1\cdots i_n}
\ea
the asymptotic equation, \eqref{eq:asym_3_decomp}, becomes
\ba
 &\f{1}{6}
  \Biggr[\tau^{a_1a_2a_3}_{r_1r_2r_3}+\f{\aop^{a_1a_2a_3}}{2}
    \rob{\f{V^{a_1a_2}_{c_2c_3}}{2E^{a_1a_2a_3}}\tau^{a_3c_2c_3}_{r_1r_2r_3}}
  +
 \cr &
 \f{\aop^{a_1a_2a_3}_{r_1r_2r_3}}{4}
 \rob{\f{V^{a_1a_2}_{r_3c_3}\tau^{c_3a_3}_{r_1r_2}}{E^{a_1a_2a_3}}}\Biggl]
  L^{r_1r_2r_3;d_4\cdots d_n}_{i_1\cdots i_n} = 0
\;.\ea
A sufficient demand for this equation to hold is 
\ba\label{eq:tau3}
  \tau^{a_1a_2a_3}_{r_1r_2r_3}
  &+\f{\aop^{a_1a_2a_3}}{2}
    \rob{\f{V^{a_1a_2}_{d_2d_3}}{2E^{a_1a_2a_3}}\tau^{a_3d_2d_3}_{r_1r_2r_3}}
  \cr & +
 \f{\aop^{a_1a_2a_3}_{r_1r_2r_3}}{4}
 \rob{\f{V^{a_1a_2}_{r_3d_3}\tau^{d_3a_3}_{r_1r_2}}{E^{a_1a_2a_3}}}
 = 0\;.
\ea
Writing it in matrix form (in the non symmetrized basis)
\ba
   \hat{\tau}_3-\q{3}\gop\rob{0}\vop\q{3}\hat{\tau}_3
   -\q{3}\gop\rob{0}\vop\hat{\tau}_2 = 0
\ea
we find the formal solution
\ba
   \hat{\tau}_3&=\f{1}{1-\q{3}\gop\rob{0}\vop}\q{3}\gop\rob{0}
   \vop\hat{\tau}_2\;.
\ea
As in the two-body case, this result is related to the asymptotics of the 3-body amplitude \cite{BWB23A}
\be
   \opt{3} \to \opty{3} 
    =
    \f{1}{1-\q{3}\gop\rob{0}\vop}\q{3}\gop\rob{0}\vop\opty{2}\p{3}
\ee
through the relation $\opty{3}=\hat{\tau}_3\p{3}$.

%
%
\section{The double and triple Coupled Cluster equations with known asymptotics}\label{sec:cc_eq_asym}

Once known, the asymptotic behaviour of the CC amplitudes can be utilized to
eliminate high momentum components from the CC equations.
In the following we outline this procedure at the level
of the coupled-cluster-doubles-triples (CCDT) approximation, 
i.e. assuming $\opt{n}=0$ for $n\geq 4$. We note that at the 2-body CCD level
this procedure is very similar to the iterative technique used to solve the CC equations.

To eliminate the asymptotic part let us define the residual 2, and 3-body operators
\ba\label{def_delta23}
   \hat{\Delta}_2&\equiv \opt{2}-\opty{2}\;,
   \cr
   \hat{\Delta}_3&\equiv \opt{3}-\opty{3}-\reb{\hat{\tau}_2,\hat{\Delta}_2}_3\;,
\ea
where the subscript $3$ denotes the fact that only the 3-body part of the commutator should be kept.

The reasoning behind the definition of $\hat\Delta_2$ is clear, eliminating the
2-body asymptotics encapsulated in $\opty{2}$.
The 3-body case is a bit more complicated. Here there are two types of asymptotics, the highly excited 2-particle case, and the 3-particle case.
The operator $\opty{3}$ clearly deals with the latter case.
The 3-body asymptotics of $\reb{\hat{\tau}_2,\hat{\Delta}_2}_3$ is given by
\ba
   \reb{\hat{\tau}_2,\hat{\Delta}_2}_3\to   
   \f{\aop^{a_1a_2a_3}_{i_1i_2i_3}}{4}&\rob{\tau^{a_1a_2}_{i_1d_4}\Delta^{d_4a_3}_{i_2i_3}}.
\ea
Since in this limit $\Delta^{d_4a_3}_{i_2i_3}\ll t^{d_4a_3}_{i_2i_3}$, and
$\tau^{a_1a_2}_{i_1d_4}\propto -\f{V^{a_1a_2}_{i_1d_4}}{E^{a_1a_2}}$
this term is negligible with respect to $\opt{3}$.

Considering now the
2-body asymptotics, we first 
utilize Eqs. \eqref{2b_fctr_ansatz} and \eqref{ll2_me} to get
the $a_1,a_2\to\infty$ limit of $\opt{3}$, 
\ba\label{T3_2b_limit}
   {t}^{a_1a_2d_3}_{i_1i_2i_3} &=
   \frac{1}{2}\tau^{a_1a_2}_{r_1r_2}L^{r_1r_2;d_3}_{i_1i_2i_3} 
   =
   \frac{1}{4}\tau^{a_1a_2}_{r_1r_2}
   \aop^{r_1r_2}_{i_1 i_2 i_3}\rob{\delta^{r_1}_{i_1}t^{r_2d_3}_{i_2i_3}}
   \cr &=
   \frac{1}{2}\aop_{i_1 i_2 i_3}\rob{\tau^{a_1a_2}_{i_1r_2}t^{r_2d_3}_{i_2i_3}}.
\ea
Exploring now the same limit for $\opty{3}$, we note that we can repeat all the arguments in section \ref{sec:2b_reduction} for $\opty{3}$, applying them
e.g. to Eq. \eqref{eq:tau3}. The resulting expression for $\opty{3}$
in the limit $a_1,a_2\to\infty$ is
\be\label{T3inf_2b_limit}
   \rob{t^\infty}^{a_1a_2d_3}_{i_1i_2i_3} =
   \frac{1}{2}\aop_{i_1 i_2 i_3}\rob{\tau^{a_1a_2}_{i_1r_2}
     \rob{t^\infty}^{r_2d_3}_{i_2i_3}}.
\ee
Taking the difference of these two equations we get
\ba\label{eq:tmtinf}
  t^{a_1a_2d_3}_{i_1i_2i_3}&-\rob{t^\infty}^{a_1a_2d_3}_{i_1i_2i_3}
  \cr&=
  \f{\aop_{i_1i_2i_3}}{2}\rob{\tau^{a_1a_2}_{i_1r_2}t^{r_2d_3}_{i_2i_3}-
    \rob{\tau^{a_1a_2}_{i_1r_2}\rob{t^\infty}^{r_2d_3}_{i_2i_3}}}
  \cr&=
  \f{\aop_{i_1i_2i_3}}{2}\rob{\tau^{a_1a_2}_{i_1r_2}\Delta^{r_2d_3}_{i_2i_3}}
  \;.
\ea
Asymptotically, this difference is canceled out completely by the commutator
\be \label{commutator3}
   \rob{\reb{\hat{\tau}_2,\hat{\Delta}_2}_3}^{a_1a_2d_3}_{i_1i_2i_3} = 
   \f{\aop^{a_1a_2d_3}_{i_1i_2i_3}}{4}
   \rob{\tau^{a_1a_2}_{i_1r_2}\Delta^{r_2d_3}_{i_2i_3}}.
\ee
To prove this claim we observe that as $a_1,a_2\to\infty$ terms such as $\Delta^{r_2a_1}_{i_1i_2}$ can be neglected. Therefore the permutation over the upper indices in \eqref{commutator3} yields a factor of 2, and
\ba\label{eq:delta_comm_asym}
   \rob{\reb{\hat{\tau}_2,\hat{\Delta}_2}_3}^{a_1a_2d_3}_{i_1i_2i_3} \approx
   \f{\aop_{i_1i_2i_3}}{2}
   \rob{\tau^{a_1a_2}_{i_1r_2}\Delta^{r_2d_3}_{i_2i_3}}.
\ea
Hence, the definition  introduced in Eq. \eqref{def_delta23} ensures that
both the 2-particle, and the 3-particle asymptotics are eliminated from $\hat\Delta_3$. 

As asymptotically both $\hat{\Delta}_2$ and $\hat{\Delta}_3$ converge to zero
much faster than the corresponding CC amplitudes $\opt{2},\opt{3}$, we expect it to be more beneficial to solve first 
the integral equations for $\hat{\tau}_2$ and $\hat{\tau}_3$
and then
the CC equations for $\hat{\Delta}_2,\hat{\Delta}_3$.
For completeness we present the CCDT equations for $\hat{\Delta}_2,\hat{\Delta}_3$,
\begin{widetext}
\ba
  0=\bra{\Phi_{i_1i_2}^{d_1d_2}}
  \vop&+[\hat{H}_{0},\opty{2}]
  + [\vop,\opty{2}]
  +\f{1}{2}[[\vop,\opty{2}],\opty{2}]
  + [\vop,\opty{3}]
  \cr &
  +[\hat{H}_{0},\hat{\Delta}_2]
  + [\vop,\hat{\Delta}_2]
  +\f{1}{2}[[\vop,\hat{\Delta}_2],\hat{\Delta}_2]
  +[[\vop,\hat{\Delta}_2],\opty{2}]
  + [\vop,\hat{\Delta}_3]
  + [\vop,[\hat{\tau}_2,\hat{\Delta}_2]_3]
  \ket{\Phi_0},
  \\
  0=\bra{\Phi_{i_1i_2i_3}^{d_1d_2d_3}}&
    [\hat{H}_{0},\opty{3}] 
  + [\vop,\opty{2}]
  + \f{1}{2}[[\vop,\opty{2}],\opty{2}]
  + [\vop,\opty{3}]
  + [[\vop,\opty{2}],\opty{3}]
  \cr&
  + [\hat{H}_{0},\hat{\Delta}_3]
  + [\vop,\hat{\Delta}_3]
  + [\hat{H}_{0},[\hat{\tau}_2,\hat{\Delta}_2]_3]
  + [\vop,\hat{\Delta}_2]
  + \f{1}{2}[[\vop,\hat{\Delta}_2],\hat{\Delta}_2]
  + [[\vop,\hat{\Delta}_2],\opty{2}]
  + [[\vop,\hat{\Delta}_2],\opty{3}]  
  \cr&
  + [\vop,[\hat{\tau}_2,\hat{\Delta}_2]_3]
  + [[\vop,\hat{\Delta}_2],\hat{\Delta}_3]+ [[\vop,\opty{2}],\hat{\Delta}_3]
  + [[\vop,\hat{\Delta}_2],[\hat{\tau}_2,\hat{\Delta}_2]_3]
  + [[\vop,\opty{2}],[\hat{\tau}_2,\hat{\Delta}_2]_3]
  \ket{\Phi_0}.
  \ea
\end{widetext}
%
%
\section{Summary}\label{sec:summary}
Summing up, considering a closed shell system of identical fermions
interacting via a 2-body potential, we have analyzed 
the asymptotic behaviour of the coupled cluster 
many-body wave-function in the limit of  highly excited two and three particles states.
Under few assumptions regarding the nature of this 2-body interaction we have found that in this limit the different coupled cluster amplitudes exhibit a recurring 
  behaviour factorizing into a product of a common asymptotic two- or 
  three-body amplitudes and a low energy term.
  The asymptotic terms depend on the potential and in general are system specific.


Utilizing these results, the asymptotic behaviour of the CC amplitudes can be eliminated.
As an example we have analyzed the $\opt{2}$ and $\opt{3}$ amplitudes
within the CCDT approximation.

Some of the limitations of the current work, e.g. the absence of 1-body and/or 3-body interaction, as well as the restriction of the analysis to closed shell systems call for further investigation and are the subject of future works.

%
%
\begin{acknowledgments}
This research was supported by the ISRAEL SCIENCE FOUNDATION 
(grant No. 1086/21).
The work of S. Beck was also supported by the Israel Ministry of Science and Technology (MOST).
R. Weiss was supported by the Laboratory Directed Research and Development program of Los Alamos National Laboratory under project number 20210763PRD1.
\end{acknowledgments}
%
%
%
\bibliography{references}
\end{document}